\documentclass[referee,useAMS,usenatbib]{mn2e}
\usepackage[utf8]{inputenc}
\usepackage{amsmath}
\usepackage{graphicx}
\usepackage{epsfig}
\usepackage{chngcntr}
\usepackage[font={small,it}]{caption}
\usepackage{textcomp}
\usepackage{subfig}

\def\ie{{\em i.e., }}
\def\vr{v_r}
\def\vfi{v_{\phi}}
\def\ra{r_{\rm \small A}}
\def\rs{r_{\rm \small S}}
\def\rf{r_{\rm \small F}}
\def\bfi{B_{\phi}}
\def\rg{r_{\rm g}}
\def\nel{n_{{\rm e}^-}}
\def\np{n_{\rm p}}
\def\mel{m_{{\rm e}^-}}
\def\mp{m_{\rm p}}
\def\us{u_{\rm \small S}}
\def\ua{u_{\rm \small A}}
\def\uf{u_{\rm \small F}}
\def\vra{v_{{\rm \small A}r}}
\def\m2a{M^2_{\rm \small A}}
\def\ma{M_{\rm \small A}}

\def\lsim{\lower.5ex\hbox{$\; \buildrel < \over \sim \;$}}
\def\gsim{\lower.5ex\hbox{$\; \buildrel > \over \sim \;$}}
\newcommand{\RN}[1]{
}

\usepackage{graphicx}
\title[]
{Effect of plasma composition on magnetized outflows}
\author[Singh \& Chattopadhyay]
{Kuldeep Singh$^{1,2}$, Indranil Chattopadhyay$^{1}$\thanks{Email:
kuldeep@aries.res.in (KS); indra@aries.res.in (IC)}\\
$^{1}$Aryabhatta Research Institute of Observational Sciences 
(ARIES), Manora Peak, Nainital-263002, India.\\
$^{2}$University of Delhi, Delhi, India.}
\begin{document}
\date{}
\maketitle
\label{firstpage}

\begin{abstract}
In this paper, we study magnetized winds described by variable adiabatic index equation of state in Paczy\'{n}ski \& Wiita
pseudo-Newtonian potential. We identify the flow solutions with the parameter space of the flow. We also confirm
that the physical wind solution is the one which passes through the slow, Alfv\'en and fast critical points. We study the dependence
of the wind solution on the Bernoulli parameter $E$ and the total angular momentum $L$. The adiabatic
index, which is a function of temperature and composition, was found to be variable in all the outflow solutions.
For the same values of the Bernoulli parameter and the total
angular momentum, a wind in strong gravity is
more accelerated, compared to a wind in Newtonian gravity. We show that flow variables like the
radial and azimuthal velocity components, temperature all depend on the composition of the flow.
Unlike the outflow solutions in hydrodynamic regime, the terminal speed of a  magnetically driven wind also
depends on the composition parameter.
\end{abstract}
\begin{keywords}
{Magneto hydrodynamics (MHD); Outflows; Stars: Neutron stars; Black Hole}
\end{keywords}

\section{Introduction}
\label{sec:intro}
There are many magneto-hydrodynamic (MHD) studies of winds around central gravitating objects. For example, 
\citet{me67, me68} studied loss of angular momentum and magnetic breaking and \citet[][hereafter WD]{wd67}
studied the solar wind by solving MHD equations self-consistently. Their model is well tested 
and predicted 
the wind speed at the earth’s orbit. Studies of winds were further carried out by \citet{pn71, ok74, ok75}. Later
\citet{sa85, sa87} generalized WD wind model and studied wind away from the equatorial plane and its collimation
by the magnetic field. The generalization of magnetized winds later became the starting point of studies on
magnetically driven jets \citep{c86,lbc91,lrb95,dd02,pmm10}. {It may be noted that, \citet{bp82} studied accretion disc particles
being flung along the poloidal magnetic field lines, the foot points of which are co-rotating with the Keplerian accretion disc.
This model is similar to a bead on a wire scenario. The assumption of cold flow limits the solutions to a trans-Alfvenic
and trans-fast flow and therefore, these solutions are oblivious of the location of slow-magnetosonic points. Moreover, one may note,
Keplerian disc may not be the base of the jet but the corona (or, the hotter, inner part of the accretion disc) might actually be the base.
Therefore, studying hot flow near the compact object is quite important.}

{The outflow models which did consider hot flow \citep[for e. g.,][]{c86,lbc91}}, the authors have used constant adiabatic index ($\Gamma$) equation of state (EoS) to describe
the thermodynamics of the plasma. However, if we implement any of 
these models at regions around compact objects (hot stars, neutron stars, black holes), which are hotter than the environ of
ordinary stars, then 
fixed $\Gamma$ EoS is not valid. In other words, close to the central object, the flow should be hot enough such that $\Gamma \gsim 4/3$,
but at asymptotically large distances $\Gamma \sim 5/3$, which is equivalent to temperature
variation of more than four orders of magnitude. Such temperature variation is expected in outflows.
In this paper, instead of focusing on jets, we concentrate on the role that correct plasma thermodynamics
may play on the WD type wind solutions.
We consider a
variable $\Gamma$ equation of state \citep[abbreviated as CR EoS,][]{cr09} and obtain outflow solutions for winds around
compact objects. We use 
Paczy\'{n}ski \& Wiita (PW) potential \citep{pw80} to study the behaviour around a stronger gravity. However,
we would like to understand the role which strong gravity might play on such wind solutions by comparing with the winds
in Newtonian potential.

One of the advantage of using CR EoS is that, this EoS 
has composition parameter $\xi$, therefore we can also study effect of composition on wind solution
in the present manuscript. The composition of the flow, controls the inertial and 
thermal energy budget of the outflow. It may be noted that, the effect of composition has been studied (by employing CR EoS) in outflows
and jets \citep{vkmc15,vc17,vc18,vc19},
as well as, in accretion \citep{kscc13,kc14,ck16,kc17}, but all in the realm of hydrodynamic regime.
CR EoS has been employed for magnetized accretion onto neutron stars \citep{sc18}, but assumption of strong magnetic field simplified the
equations of motion immensely such that the flow remained sub-Alfv\'enic. Therefore, this will probably be the first study
where the issue of composition can be systematically addressed for flows in the proper MHD scenario.
In this paper, we would like to investigate the entire parameter space
for various flow parameters including the composition parameter of the flow. We would also like to investigate 
how the actual flow solution is affected by the composition of the flow. We intend to investigate many such questions
in this paper. 

The present manuscript is arranged in the following way.
In the section \ref{subsec:govereqs}, we present 
ideal magneto-hydrodynamic (MHD) equations. In the section \ref{subsec:eos}, we discuss
the relativistic equation of state (EoS) having temperature dependent adiabatic index.
In the section \ref{sec:meth}, we explain the methodology to solve the equations of motion.
In the section \ref{sec:result}, we present the parameter space for the critical points and
outflow/wind solutions. Finally in section \ref{sec:conclude} we present discussions and the concluding remarks.

\section{MHD equations and assumptions}
\subsection{Governing equations}
\label{subsec:govereqs}
We assume that the flow is steady, inviscid and a highly conducting plasma. Therefore, MHD equations have the
following form \citep{wd67, h78},
\begin{equation}
\nabla\ldotp(\rho \textbf{v}) = 0,
\label{cont.eq}
\end{equation}
\begin{equation}
\nabla\ldotp\textbf{B} = 0,
\label{nomonopol.eq}
\end{equation}
\begin{equation}
\nabla\times(\textbf{v}\times \textbf{B}) = 0,
\label{faraday.eq}
\end{equation}
\begin{equation}
(\rho \textbf{v}\ldotp\nabla)\textbf{v} = - \nabla p + \frac{1}{c}(\textbf{J}\times\textbf{B}) - \Phi^{\prime}(r)\textbf{\^{r}}
\label{mblnc.eq}
\end{equation}
Here, $\Phi(r)$ is the gravitational potential and 
$\Phi(r) = \Phi_{\rm NP}(r) = -GM/r$, is the Newtonian potential and its derivative is 
$\Phi^\prime(r) = \Phi^\prime_{\rm NP}(r) = GM/r^2$. The PW potential
$\Phi(r) = \Phi_{\rm PWP}(r) = -GM/(r - \rg)$ and its derivative is
$\Phi^\prime(r) = \Phi^\prime_{\rm PWP}(r) = GM/(r - 
\rg)^2$ where the Schwarzschild radius is $\rg=2GM/c^2$, $G$ is the gravitational constant and $M$ is the mass of the central object
and $c$ is the speed of light in vacuum.
Assuming ideal MHD, we integrate MHD equations along magnetic field lines and axis symmetry assumption to
obtain the conserved quantities as:\\
(i) The mass flux conservation is obtained from the continuity equation (\ref{cont.eq}),
\begin{equation}
 \rho v_{r} r^2 = {\rm constant} = \dot{M},
 \label{conMp.eq}
\end{equation}
(ii) The magnetic flux conservation is obtained from the Maxwell's equation (\ref{nomonopol.eq}),
\begin{equation}
 B_{r}r^2 = {\rm constant} = B_{\circ}r^{2}_{\circ},
 \label{conBFp.eq}
\end{equation}
(iii) The Faraday equation (\ref{faraday.eq}), for highly conducting fluid gives, 
\begin{equation}
 r(v_{r}B_{\phi} - v_{\phi}B_{r}) = {\rm constant} = - \Omega B_{r}r^2 ,
 \label{conFp.eq}
\end{equation}
(iv) $r^{th}$ component of momentum balance equation (\ref{mblnc.eq}) gives the total energy conservation,
\begin{equation}
\frac{1}{2}v_{r}^2 + \frac{1}{2}v^{2}_{\phi} + h + \Phi(r) - \frac{B_{\phi}B_{r}\Omega r}{4\pi\rho v_{r}}
 = {\rm constant} = E,
 \label{conEngp.eq}
\end{equation}
(v) $\phi^{th}$ component of momentum balance equation (\ref{mblnc.eq}) gives the total angular momentum conservation,
\begin{equation}
rv_{\phi} - \frac{B_{\phi}B_{r} r}{4\pi\rho v_{r}} = {\rm constant} = L.
\label{conAngp.eq}
\end{equation}
Here $r$ is the radial distance, $r_{\circ}$ is the radius of a star, or, the radial distance near black hole,
$\rho$ is the mass density, $v_{r}$ is the radial
velocity component, $v_{\phi}$ is the azimuthal
velocity component, $B_{r}$ is the radial magnetic field and subscript `$\circ$' denote the magnetic
field at distance $r_{\circ}$, $B_{\phi}$ is the azimuthal magnetic field,
$\Omega$ is the angular velocity of star or matter at $r_{\circ}$. In equation (\ref{conAngp.eq}), we see that total angular momentum has two
terms, the first term is the angular momentum associated with matter and the second term
represents the angular momentum associated with the magnetic field. Therefore, only sum of
both angular momenta is conserved and not the individual entities. This also imply that angular momentum can be exchanged
between matter and field. The radial Alfv\'{e}nic Mach number is defined by
\begin{equation}
\m2a = \frac{4\pi\rho v_{r}^{2}}{B^{2}_{r}},
\label{mach1.eq}
\end{equation}
From equations (\ref{conFp.eq}) and (\ref{conAngp.eq}), we can derive the expression for $\vfi$, 
\begin{equation}
v_{\phi} = \Omega r\frac{(\m2a Lr^{-2}\Omega^{-1}-1)}{(M^{2}_{\rm \small A}-1)}.
\label{vfip.eq}
\end{equation}

\subsection{Variable $\Gamma$ EoS}
\label{subsec:eos} 
So far we have discussed that adiabatic
index does not remain constant throughout the flow. \citet{c38} obtained the exact EoS of
hot gas which has a variable adiabatic index. But it is difficult to use this EoS in numerical calculations
due to the presence of modified Bessel functions of second kind, which we know has a non terminating form.
However, there is
another approximate but accurate EoS (compared to Chandrasekhar EoS) given by \citet{cr09} for multi-species flow (\ie composed
of electron, positron
and proton) having variable adiabatic index --- the CR EoS. In our analysis, we use CR EoS, because it has a simple
functional form $f(\Theta, \xi)$ instead of complicated Bessel functions. 
The energy density $(\bar{e})$ is given by,
\begin{equation}
\begin{split}
\bar{e} = & \frac{\rho c^{2}f(\Theta,\xi)}{K};\mbox{ where, } \\
& f(\Theta,\xi) = (2-\xi)\left[1 + \Theta \frac{(9\Theta + 3)}{
(3\Theta + 2)}\right] + \xi\left[\frac{1}{\eta}
+ \Theta \frac{(9\Theta + 3/\eta)}{(3\Theta + 2/\eta)}\right].
\label{etrnl.eq}
\end{split}
\end{equation}
The $\Theta=\kappa_BT/(\mel c^2)$ is proportional to $T$, $\rho=\nel \mel K$ is the total rest-mass 
density, $\eta=\mel/\mp$ is electron to proton mass ratio and $\xi=\np/\nel$
is the composition parameter which is the ratio of number density of protons to the number density of 
electrons. The constant $K = [2-\xi(1-1/\eta)]$
depends on the composition of the flow. A plasma described by $\xi=0.0$ has no protons and therefore is an electron-positron plasma,
for $0.0<\xi<1.0$ we have electron-positron-proton plasma and $\xi=1.0$ signify electron-proton plasma. 
{It may be noted that, $\xi=0.0$ plasma is unlikely to exist as a global solution i. e., spanning from a region close to
the compact object to asymptotically large distance. As a result, no such solution has been presented in the results
section (\ref{sec:result}). We expect the composition of the outflow to be electron-proton, however for
flows which start with very high temperature or in intense high energy radiation region (near black hole or neutron star)
may have significant electron-positron
pairs in the flow, so pair dominated plasma (with some protons) is quite possible. Since
composition of outflows have not been conclusively established in observations,
therefore in this paper, we have studied how composition of the plasma may affect the outflow solutions by using $\xi$
as a parameter in the range $0<\xi\leq 1$. 
However, to predict an observational signature of composition would require detailed radiative processes of magnetized plasma, which is
beyond the scope of the present work}.

Enthalpy $(h)$, variable adiabatic index $(\Gamma)$, polytropic index $N$ and sound speed $(c_{\rm s})$ are given by,
\begin{equation}
h = \frac{\bar{e} + p}{\rho} = \frac{fc^{2}}{K} + \frac{2\Theta c^{2}}{K}
\label{enthp.eq}
\end{equation} and 
\begin{equation}
\Gamma = 1 + \frac{1}{N},~
N = \frac{1}{2}\frac{df}{d\Theta} \mbox{ and } c^{2}_{\rm s}=\frac{2\Theta \Gamma c^{2}}{K}.
{\label{gama.eq}}
\end{equation}
The adiabatic equation of state can be obtained by integrating the $1^{\rm st}$ law of thermodynamics with
the help of the continuity equation and CR EoS \citep{kscc13, vkmc15, sc18}, to obtain
\begin{equation}{\label{rel_rho.eq}}
\rho={\cal Q}\mbox{exp}(k_3) \Theta^{3/2}(3\Theta+2)^{k_1}(3\Theta+2/\eta)^{k_2},
\end{equation}
where, $k_1=3(2-\xi)/4$, $k_2=3\xi/4$ and $k_3=(f-K)/(2\Theta)$ and ${\cal Q}$ is the measure of the
entropy. Using equation (\ref{conMp.eq}), the entropy-accretion rate $\dot{\cal{M}}$ is given by,
\begin{equation}{\label{entro.eq}}
\dot{\cal{M}}=\frac{\dot M}{4\pi {\cal Q}}=v_{r}r^{2}\mbox{exp}(k_3) \Theta^{3/2}(3\Theta+2)^{k_1}(3\Theta+2/\eta)^{k_2}
\end{equation}
One may note that, ${\dot \cal M}$ is a temperature and composition dependent measure of entropy, which remains
constant along a non-dissipative flow, or in absence of shocks.

\section{Methodology}
\label{sec:meth}
We know that plasma has three signal speeds \ie slow speed $(\us)$, Alfv\'en speed (in our case we are using radial Alfv\'en speed $\ua$) and
fast speed $(\uf)$. In the present case, the order of these speeds are $\us < \ua < \uf$. We know that
for outflows, radial velocity ($\vr$) is very small near the surface of star or a radius near the black hole,
therefore, $\ma \ll 1$ and very far from the central object, $\ma \gg 1$.
Therefore, at certain radius (say $\ra$), $v_r$ is equal to $\ua$
(\ie $v_{r}|_{\ra} \equiv \vra = \ua$) but at that radius, denominator of $\vfi$
is zero (see equation \ref{vfip.eq}). Thus the numerator should also be zero at that critical radius
to make $\vfi$ always finite and this point is known as Alfv\'{e}nic critical point.
Therefore, numerator of $\vfi$  gives a relation between the critical radius
of the Alfv\'{e}nic point and total angular momentum,
\begin{equation}
L = \Omega r^{2}_{\rm \small A}.
\end{equation}
Using equations (\ref{conMp.eq}), (\ref{conBFp.eq}) and (\ref{mach1.eq}), we can 
also write $\m2a$ as,
\begin{equation}
\m2a = \frac{\vr r^2}{\vra r^{2}_{\rm \small A}} = \frac{\rho_{\small \rm A}}{\rho}.
\end{equation}
and $\vfi$ and $\bfi$ become, 
\begin{equation}
\vfi = \frac{\Omega r}{\vra} \left(\frac{\vra-v_{r}}{1-\m2a}\right) ~~\mbox{and}~~
\bfi = -B_{r}\frac{\Omega r}{\vra \ra^{2}} \left(\frac{\ra^2-r^{2}}{1-\m2a}\right).
\end{equation} 

The $r^{th}$-component of momentum balance equation (\ref{mblnc.eq}) gives the equation of motion,
\begin{equation}
\frac{dv_{r}}{dr} = \frac{\cal N}{\cal D}, 
\label{eom.eq}
\end{equation}
\begin{equation}
{\cal N} = {\vr r}\left\{\left(\frac{2c^{2}_{s}}{r^2} - \frac{\Phi^{'}(r)}{r}\right)(\m2a-1)^3 + \Omega^2 \left(\frac{\vr}{\vra}-1\right)\left[(\m2a+1)\frac{\vr}{\vra}-3\m2a+1\right]\right\}\nonumber
\end{equation}
\begin{equation}
{\cal D} = \left(\vr^2-c^{2}_{s}\right)(\m2a-1)^{3}-\frac{\Omega^{2}}{r^{2}}\ma^4 \left(r^{2}-\ra^2\right)^{2}\nonumber.
\end{equation}

To find wind solution, we need two input parameters \ie $E$, $L$, initial boundary conditions
and composition of flow $\xi$ which is a free parameter. In our case, we use Alfv\'en point $\ra$ as the initial 
condition because at that radius, equation(\ref{eom.eq}) or $d\vr/dr \rightarrow 0/0$. Thus, equating the numerator and denominator
to zero, provides us with the critical point conditions and hence acts as mathematical boundary
conditions.
However, $d\vr/dr \rightarrow 0/0$ at other critical points when $\vr=\us$ 
or $\vr=\uf$. These critical points are known as the slow critical point ($\rs$) or the fast critical point ($\rf$), respectively.
Therefore, ${\cal N}=0$ and ${\cal D}=0$ are the critical point conditions to find all the critical points (slow points, Alfv\'en
 points, fast points and all of which can either be X-type or O-type) for a given set of input parameters.
We have found that for a small energy range and given angular momentum, there exists possibility of five critical points.
By supplying $L$, $\ra$ and $\vra$, from the critical point conditions we can find out critical radius ($r_c$) and critical
radial velocity $v_{rc}$.
Then, the total energy $E=E_{c}$ at the critical point can be calculated from equation (\ref{conEngp.eq})
and entropy accretion rate ($\dot{\cal M}=\dot{\cal M}_{c}$) from equation (\ref{entro.eq}). With the help of L'Hospital rule,
we determine the gradient of $\vr$ at the critical points.
We integrate the equation of motion forward and backward from the critical points
with the help of $4^{\rm th}$ order Runge-Kutta method and find the complete wind solution.

\vspace{0.0cm}
\begin{figure}
\hspace{0.0cm}
\includegraphics[width=16cm,height=8cm]{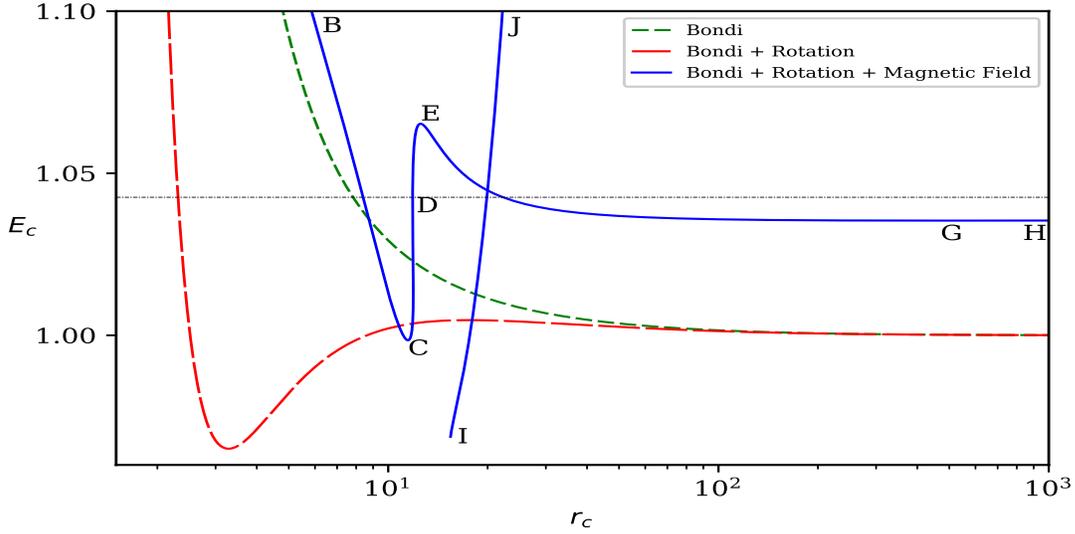}
\caption{{In this plot, total energy ($E_c$) is plotted versus critical radius ($r_c$) for Bondi flow
(dashed-green), Bondi flow with rotation (long-dashed-red) and magnetized, rotating flow (solid-blue) for $L=1.75$. Thin dashed line represent E=1.04257.}}
\label{lab:fig1}
\end{figure}

\section{Results}
\label{sec:result} 
In this paper, the unit of velocity is the speed of light $c$ in vacuum and that of distance is the Schwarzschild radius $\rg$.
We have used Alfv\'en point as initial conditions, so we have chosen
$\ra=11.85$ and $\vra=0.167$ 
for Fig.(\ref{lab:fig1}), where we have plotted Bernoulli parameter or specific energy ($E_{c}$) at the critical point versus
the critical radius ($r_{c}$) for Bondi flow (dashed, green), Bondi flow with
rotation (long-dashed, red) and Bondi flow with both the rotation and magnetic field (solid, blue)
for total angular momentum $L=1.75$ and $\xi=1$ where strong gravity is mimicked by PW potential.
Purely Bondi flow (i. e., hydrodynamic flow and only radial velocity component) harbours a single critical point
(a sonic point where $\vr=c_s$) for any value of $r_c$, {which is clear from the dashed (green) curve, which is a
monotonically decreasing function of $r$. It may be noted, that the sonic or critical
point occurs in a fluid, due to the presence of gravity. A sufficiently hot gas confined very close to the central object would expand against
gravity. However, due to $\sim (r-\rg)^{-2}$ nature of gravity, it will fall faster than the thermal term ($2c_s^2/r$). Since the kinetic term
gains at the expense of both gravity and thermal term, at some point $\vr \ge c_s$, i. e, the flow becomes transonic at the critical
point. The $E_c$---$r_c$ curve (dashed, green) is a monotonically decreasing function, and therefore, for any given $E=E_c$ the Bondi flow
admits only one sonic point.}

{However, for a} Bondi-rotating flow (hydrodynamic, $\vr$ and $\vfi$ components), {the effective gravity deviates from
its usual $\sim (r-\rg)^{-2}$ form due to the presence of the centrifugal term $\vfi^2/r$.
This} interplay of
rotation and gravity produces multiple sonic/critical points, {which is also} clear from the $E_c$---$r_c$ curve (long-dashed, red))
which has a maximum and minimum {for a given value of $L$. Therefore, for any value of $E=E_c$
within the two extrema, the flow would harbour multiple critical points}.

Hydrodynamics is relatively simple, since in this regime there is only one signal speed i. e., the sound speed ($c_s$)
{which is basically the propagating pressure perturbations and are isotropic in nature}.
Magnetized flow (solid-blue) is entirely a different ball game.
As has been mentioned above, there are three signal speeds in a magnetized plasma \ie
slow speed, Alfv\'en speed and fast speed. {It may be noted that propagation of the perturbations of the magnetic field is the
Alfv\'en wave, but the competition between magnetic and plasma pressure gives rise to the slow and fast magnetosonic waves.
When the plasma pressure and magnetic pressure works in phase, it is fast wave, if not then it is slow. This is precisely
the reason we have three signal speed in a magnetized plasma. Even the nature of these three waves are different, Alfv\'en is a transverse
wave, while fast and slow waves are longitudinal. Moreover, while Alfv\'en and slow waves are not isotropic, but fast wave is
quasi-isotropic. Hence, instead of sonic points ($\vr=c_s$) as we find in hydrodynamic regime, for magnetized plasma
we have slow ($\vr=\us$) and fast ($\vr=\uf$) magnetosonic points, and Alfv\'en point ($\vr=\ua$). }

{Above, we have discussed that multiple critical points may arise in non-magnetized plasma because of the presence of
angular momentum.
This also applies in magnetized plasma too. However, the additional feature is the plasma angular momentum is itself modified
by the magnetic field (equation \ref{conAngp.eq}), therefore the effective gravity is modified by the plasma angular momentum as well
as the magnetic field components.} In other words, addition of rotation in magnetized plasma leads to the
existence of one to four critical points in general, but within a small energy range ($E_{c}$) we have found five critical points.
It means the
flow can either be super-slow ($\vr>\us$) and/or super-Alfv\'en ($\vr>\ua$) and/or
super-fast ($\vr>\uf$) or we can say that flow can pass through one critical point or
through multiple critical points similar to WD wind solution \citep{wd67}. 
Curve marked BC represent X-type slow points, CD represent O-type slow points, DE are O-type fast points.
Points on the curve EG are X-type fast points, while GH curve are O-type fast points. Curve IJ is another set
of X-type slow points. The thin, dashed line is for $E=1.04257$, which represents the outflow solution in
the next figure, passes through slow, Alf\'ven, and fast points, or is equivalent to the classical WD solution. 

\begin{figure}
\hspace{0.0cm}
\includegraphics[width=16cm,height=8cm]{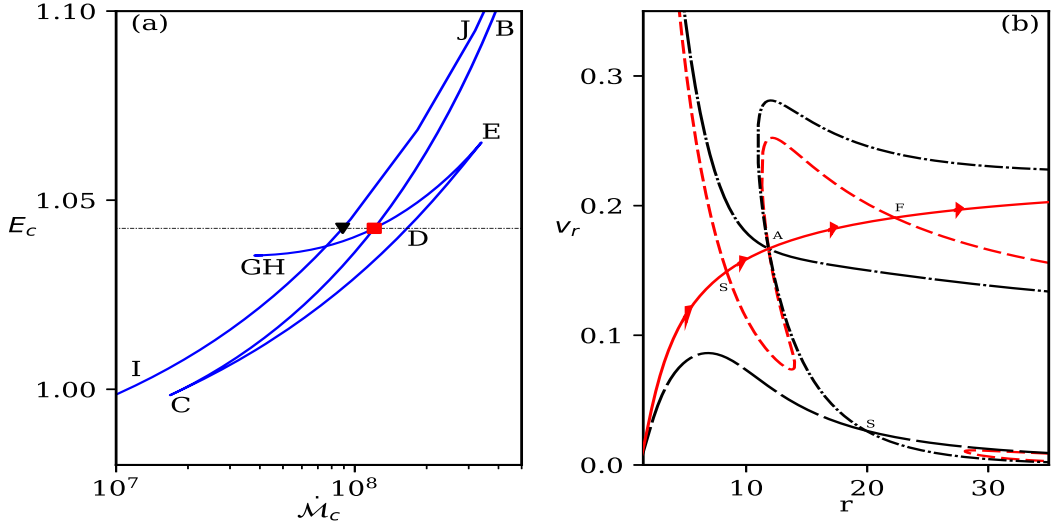}
\caption{(a) $E_c$---${\dot {\cal M}}_c$ plot corresponding to Fig. (\ref{lab:fig1}). The branches named BC,
CD etc up to IJ are also marked on the curve. The square (red) corresponds to the outflow solution
passing through slow, Alf\'ven and fast points. The triangle (black) represent solutions which is not passing through all three critical points.
(b) $\vr$ versus $r$ or actual outflow solutions. The solid (red) curve with arrow heads represent the transonic wind solution passing
through slow, Alf\'ven and fast points, marked as S, A, F, respectively. The dashed (red) curves are the transonic outflow solutions with wrong boundary condition. Long dashed curve (black) is a trans-slow solution.
Dashed-dotted curve is the trans-slow, trans-Alf\'ven solutions and long-dashed-dotted curve is a trans-Alf\'venic flow.
$E=1.04257$ (dotted, horizontal line), $L=1.75$ for all the curves.}
\label{lab:fig2}
\end{figure}

In Fig. (\ref{lab:fig2}a), we plot the $E_c$ with ${\dot {\cal M}}_c$ corresponding to the curves of Fig. (\ref{lab:fig1}).
The zones marked BC, CD, DE, EG, and GH are shown on this figure too. The parameters corresponding to the solid box (red)
corresponds to the outflow solution which passes through the slow,
Alf\'ven and fast points (i. e, the outflow is trans-slow, trans-Alf\'ven, trans-fast). {Figure (\ref{lab:fig2}a)
clearly shows that, wind represented by the solid box (red), possesses the same entropy (${\dot {\cal M}}_c$) and energy $E_c$
in all the three critical points.} The solid inverted triangle represents a flow
which has the same specific energy and total angular momentum
as the flow represented by the solid box, but passes only through the slow and/or Alfv\'en critical points and has lower entropy.
One may remember, that  to find the solutions, we have to supply $\ra$ along with $E$, $L$ and $\xi$, and obtain the value of $\vra$ by iteration.
We choose $\ra=11.85$ for all the solutions in this paper, till mentioned otherwise.
In Fig. (\ref{lab:fig2}b), we plot the actual outflow solutions, corresponding to the parameters of the
solid square of Fig. (\ref{lab:fig2}a). It may be noted $\vra=0.167$ in this case. The solid (red) curve with arrows
shows the outflow solution passing through the slow (trans-slow), Alf\'ven (trans-Alf\'ven) and fast points
(trans-fast) represented on the figure as S (i. e., $\rs$), A (i. e., $\ra$), 
and F (i. e., $\rf$), respectively. This solution passes through all the critical points and is a global solution (connecting
outflows near the compact object with asymptotically large distance). {Since the entropy of the all the critical points
are same (solid square in Fig. \ref{lab:fig2}a), the wind solution is smooth.} Among all possible global solutions,
the one passing through S, A and F has higher entropy and therefore is the correct physical solution.
This is equivalent to the WD class of solutions. The dashed curve represents the solution which also passes through S, A and F points
but with boundary conditions which are opposite to that of the outflow solution and is multi-valued in a limited range of $r$.
The boundary condition of an outflow is that it has to be sub-slow (i.e., $\vr \sim$ small)
near the central object and super-fast (i.e, $\vr \sim$ high) at asymptotically large distances, which is evidently
not the case for the dashed (red) curve. Other solutions which are
not marked with arrows, also do not satisfy the boundary conditions of an outflow.
These solutions either pass through 
A (long-dashed-dotted), or through A and S (dashed-dot), or at times only through S (long dashed). 
It is interesting to inquire about solutions for $E > 1.04257$ mark (above dotted horizontal line in Fig. \ref{lab:fig2}a).
Such solutions may have entropy higher than that corresponding to
the solid square, but unfortunately, those solutions do not pass simultaneously through $\rs,~\ra,~\&~\rf$. Moreover, if there
exist a global solution for such parameters, still we cannot consider them as proper solutions because those are decelerating.
The resulting terminal speed therefore,
is less than the solution represented by solid curve with arrows in Fig. (\ref{lab:fig2}b).
In other words, the outflow solution for any set of $E$---$L$, which passes simultaneously through $\rs$, $\ra$
and $\rf$ (one with arrows) is the correct and accelerating class of wind
solutions and was first pointed out by \citet{wd67}. 

\vspace{0.0cm}
\begin{figure}
\hspace{0.0cm}
\includegraphics[width=16cm,height=10cm]{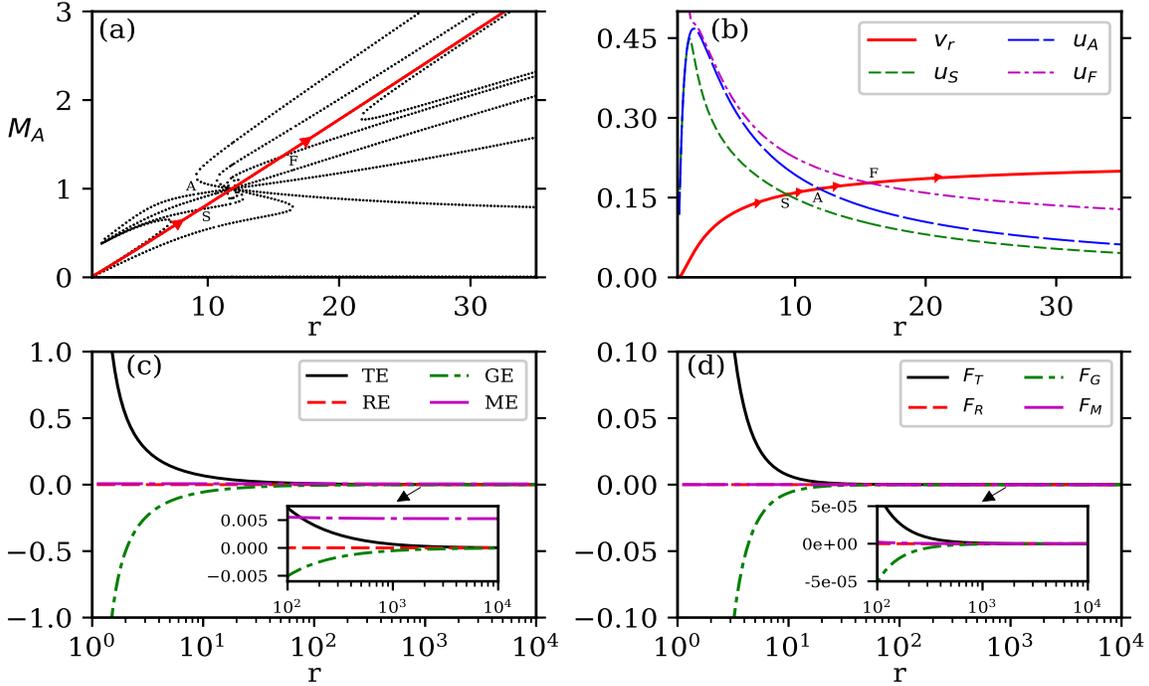}
\caption{(a) {$\ma$ vs $r$ as a function of $r$. 
The solid curve with arrow heads is the physical wind solution and the dotted are other unphysical branches.
Location of the slow magneto-sonic, Alfv\'en and fast magneto-sonic points
are marked as S, A and F, respectively.
(b) Comparing $\vr$ (solid with arrows, red), $\us$ (dashed, green), $\ua$ (long-dashed, blue) and $\uf$
(dashed-dot, violet) of the physical wind branch as a function of $r$. The locations where $\vr$ crosses $\us$, $\ua$ and $\uf$ are marked as S, A and F.
(c) Thermal (solid, black), rotational (dashed, red), gravitational (dashed-dot, green) and magnetic (long-dashed-dotted, violet)
terms of the Bernoulli parameter $E$ are named as TE, RE, GE and ME, respectively. (d) Comparison of forces ${\rm F}_{\rm T}$ (solid, black),
${\rm F}_{\rm R}$ (dashed, red), ${\rm F}_{\rm G}$ (dashed-dot, green), ${\rm F}_{\rm M}$ (long-dashed, violet) as function of $r$.
Panels (b--- d) presents variables corresponding to the solution (solid) in panel a.}
The wind is for $E=1.03075$ and $L=1.0$.
Here $\xi=1$.}
\label{lab:fig3}
\end{figure}

In Fig. (\ref{lab:fig3}a), {we plot the Alfv\'en Mach number $\ma$ (equation \ref{mach1.eq}) as a function of $r$,
corresponding to the
physical solution (trans-slow, trans-Alf\'ven,
trans-fast) represented as solid curve, marked with arrows. Physical wind solution passes simultaneously through all the three
critical points S, A and F. The dotted curves represent other unphysical solutions. It may be noted that, ordinary Mach number
distribution (i. e., $\vr/c_s$) is not presented in the figure.
This is simply because in a magnetized plasma, sound speed works in tandem with magnetic pressure and by itself is not vitally important.
However, Alfv\'en
speed determines the flow structure, therefore, the information whether a flow is super or sub Alfv\'enic is important}.
In Fig. (\ref{lab:fig3}b), we compare $\vr$ (solid with arrows, red), $\us$ (dashed, green), $\ua$ (long-dashed, blue) and $\uf$
(dashed-dot, violet) {of the physical wind solution presented in the previous panel}, as a function of $r$. These solutions corresponds to $E=1.03075$, $L=1.0$ and $\xi=1$.
The locations of $\rs$, $\ra$ and $\rf$ (marked as S, A and F in the figure) corresponds to the intersection of $\vr$ with
$\us$, $\ua$ and $\uf$. In Fig. (\ref{lab:fig3}c), we plot various components of $E$, namely, the thermal or TE ($\equiv h-1$),
the rotational or RE ($\equiv \vfi^2/2$), the gravitational or GE ($\equiv \Phi$) and the magnetic or ME
($\equiv - \{B_{\phi}B_{r}\Omega r\}/\{4\pi\rho v_{r}\}$) terms of equation \ref{conEngp.eq}. The inset zooms all the curves for
$r\rightarrow$ large. Near the compact object, TE term dominates, while at $r\gsim 100\rg$ the ME term dominates.
{Similarly, we plot
various force terms $F_{\rm T}$ (thermal), $F_{\rm R}$ (rotational), $F_{\rm G}$ (gravitational) and $F_{\rm M}$ (magnetic), along the
streamline in Fig.
\ref{lab:fig3}d. Near the central object, $F_{\rm T}$ drives the flow against gravity. At large distance
all the forces become comparable to each other, therefore comparing the combination of forces which competes with each other
gives a better picture.
The thermal force is the primary agency which opposes gravity, while magnetic force reduces $\vfi$. So we paired the
competing forces like $F_{\rm T}$ and $F_{\rm G}$ and compared with the other combination $F_{\rm R}$ and $F_{\rm M}$.
At large distance the magnetic and the centrifugal
forces together exceeds the thermal and the gravitational forces and drives the wind outward (see Fig. \ref{lab:figapend} in
Appendix \ref{app:appA}).}

\vspace{0.0cm}
\begin{figure}
\hspace{0.0cm}
\includegraphics[width=16cm,height=10cm]{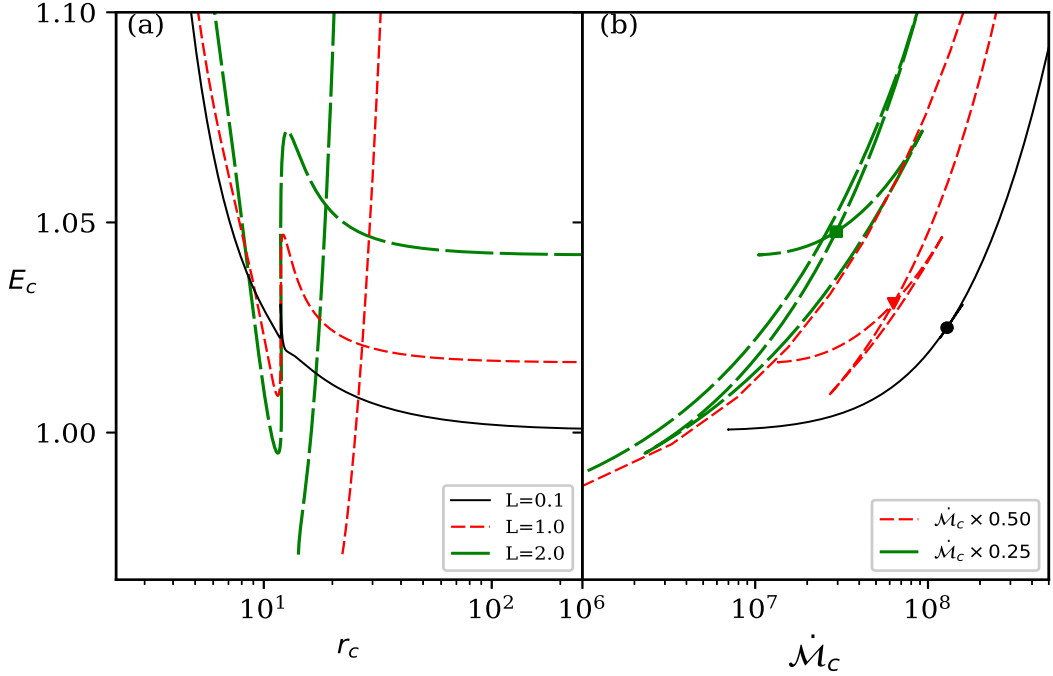} 
\caption{(a) We have plotted the total energy $E_{c}$ at critical point $r_{c}$ for total angular momentum $L=0.1$
(solid, black), $1.0$ (dashed, red)and $2.0$ (long-dashed, green).
(b) $E_{c}$---$\dot{\cal M}_{c}$ for various values of $L=0.1$
(solid, black), $1.0$ (dashed, red)and $2.0$ (long-dashed, green).
Here $\xi=1$.}
\label{lab:fig4}
\end{figure}

We study the effect of $L$ on outflow solutions. In Fig. (\ref{lab:fig4}a) we plot $E_c$ as a function of $r_c$, each curve is for $L=0.1$ (solid, black), $1.0$
(dashed, red) and $2.0$ (long-dashed, green). 
With the increase of $L$, the
flow becomes more energetic at a given critical point. In Fig. (\ref{lab:fig4}b) we plot $E_c$ versus ${\dot {\cal M}}_c$.
For each value of $L$, all the branches for O-type and X-type critical points are present, however for low $L$
(solid, black) the kite-tail part is very small, which implies that multiple critical points are possible only for a rotating flow.
Although $L$ has a significant effect on the parameter-space,
but in order to get a more quantitative idea, one need to compare outflow solutions for different $L$ but
same $E$.

\vspace{0.0cm}
\begin{figure}
\hspace{0.0cm}
\includegraphics[width=16cm,height=10cm]{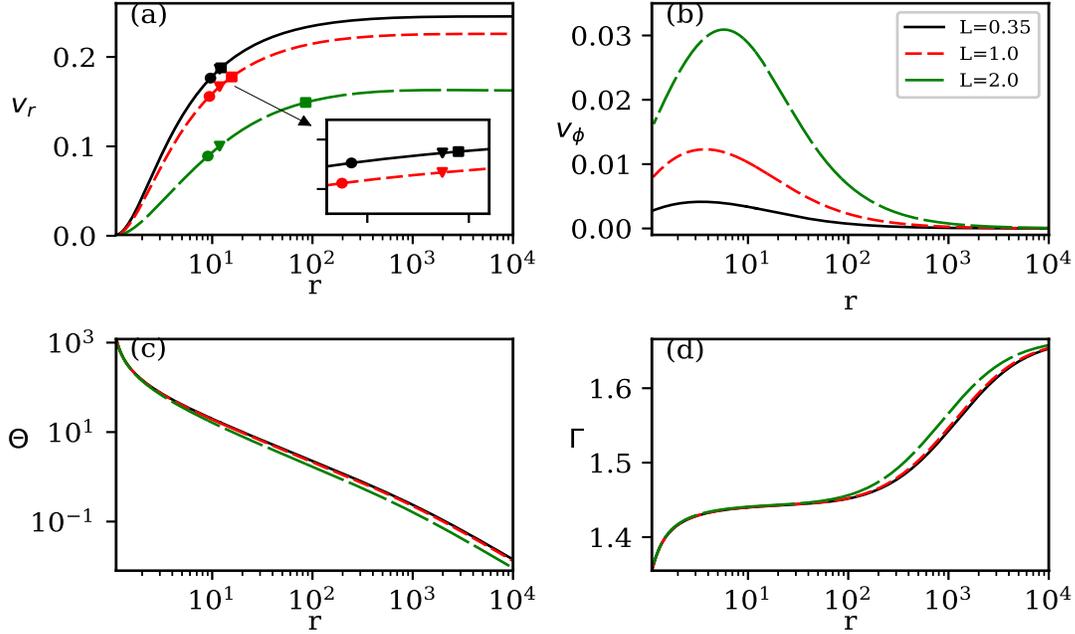} 
\caption{Flow solutions like, (a) $\vr$, (b) $\vfi$, (c) $\Theta$ and (d) $\Gamma$ as a function of $r$.
Each curve is for $L=0.35$ (solid, black), $L=1.0$ (dashed, red), $L=2.0$ (long-dashed, green) 
Here $E=1.03075$, $\xi=1$.}
\label{lab:fig5}
\end{figure}

We compare the flow solutions, like $\vr$ (Fig. \ref{lab:fig5}a), $\vfi$ (Fig. \ref{lab:fig5}b),
$\Theta$ (Fig. \ref{lab:fig5}c) and $\Gamma$ (Fig. \ref{lab:fig5}d) as a function of $r$, for various values of
$L=0.35$ (solid, black), $L=1.0$ (dashed, red), $L=2.0$ (long-dashed, green). All the plots have the same Bernoulli parameter $E=1.03075$
and $\xi=1.0$. In Fig. (\ref{lab:fig5}a), the solid circle, arrow head, and square
represents the positions of the slow, Alfv\'en and fast points, respectively. For $L=0.35$, the Alfv\'en
and fast points are almost merged, the inset zooms the region to resolve those two points.  
Outflows with higher $L$ and same $E$ have higher values of $\vfi$. 
Interestingly, outflows with higher values of $L$, are slower (low $\vr$ in long-dashed curve).
If one compares various terms in the equation (\ref{conEngp.eq}), by keeping $E$ constant but increasing $L$, then the budget in
centrifugal and magnetic terms are larger compared to that in radial kinetic and thermal terms. Therefore, $\vfi$ increases with $L$,
but $\vr$ and $\Theta$ decreases.
\vspace{0.0cm}
\begin{figure}
\hspace{0.0cm}
\includegraphics[width=16cm,height=10cm]{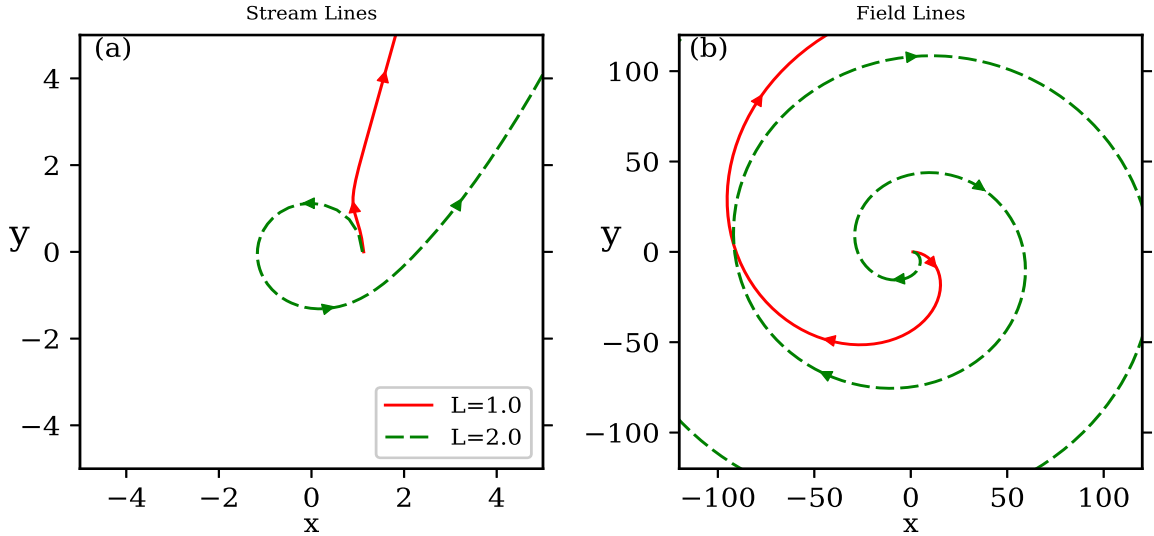} 
\caption{{(a) Flow streamlines and (b) magnetic field lines. Each of the curves are for total angular momentum
$L=1.0$ (solid, red) and
$L=2.0$ (dashed, green)}. In both the cases $E=1.03075$.}
\label{lab:fig6}
\end{figure}

{We plot the outflow streamlines in Fig. (\ref{lab:fig6}a) and the magnetic field lines Fig. (\ref{lab:fig6}b)
of a plasma
whose $E=1.03075$, where total angular momenta are $L=1.0$ (solid, red) and $L=2.0$ (dashed, green).
The wind streamline (SL) and magnetic field lines (FL)
are obtained by integrating the following equations,
\begin{equation}
d\phi_{\rm \small SL}=\frac{\vfi}{\vr}\frac{dr}{r};~\&~ d\phi_{\rm \small FL}=\frac{B_\phi}{B_r}\frac{dr}{r}
\end{equation}
This plot reconfirms the governing equations
which showed that the magnetic field is clockwise but the wind is counter clockwise. It also shows a very important effect that
magnetic field has on ionized plasma. It modifies the plasma streamline by redistributing the plasma angular momentum
($r\vfi$), such that, for $L=1.0$ flow (solid, red; Fig \ref{lab:fig6}a) which was launched with a counter-clockwise rotation, is flung out even
before completing a loop. }
So this figure presents
the structure of the outflow for two angular momenta as depicted in Figs. (\ref{lab:fig5}).
It is clear that the outflow and the magnetic field are not parallel to each other.

\vspace{0.0cm}
\begin{figure}
\hspace{0.0cm}
\includegraphics[width=16cm,height=10cm]{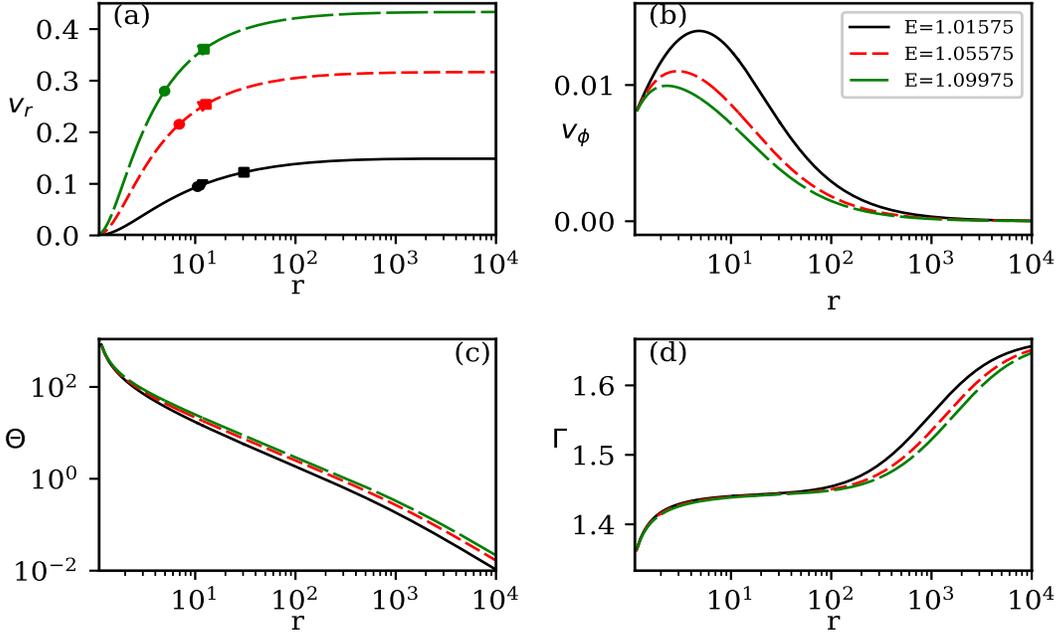} 
\caption{{These are the wind solutions for different Bernoulli parameters $E=1.01575$ (solid-black),
$1.05575$ (dashed-red)and $1.09975$ (long-dashed-green).
We have plotted the radial velocity $\vr$ (a), $\vfi$ (b), $\Theta$ (c) and
$\Gamma$ (d) versus radius $r$. All the plots are for $L=1.0$ and $\xi=1.0$.}}
\label{lab:fig7}
\end{figure}

To study the effect of Bernoulli parameter, we compare $\vr$ (Fig. \ref{lab:fig7}a), $\vfi$ (Fig. \ref{lab:fig7}b),
$\Theta$ (Fig. \ref{lab:fig7}c) and $\Gamma$ (Fig. \ref{lab:fig7}d), for various values of $E=1.01575$ (solid-black), $1.05575$ (dashed-red)and $1.09975$ (long-dashed-green) for a given value of $L=1.0$.
The solid circle, arrow head, and square
represents the positions of the slow, Alfvén and fast points, respectively.
The flow with higher $E$ is faster (high $\vr$), less rotating (low $\vfi$) and hotter (high $\Theta$) compared to flows with lower
values of $E$.
$\Gamma$ is not constant in any of the cases discussed so far and it follows the $\Theta$ distribution.

\vspace{0.0cm}
\begin{figure}
\hspace{0.0cm}
\includegraphics[width=16cm,height=10cm]{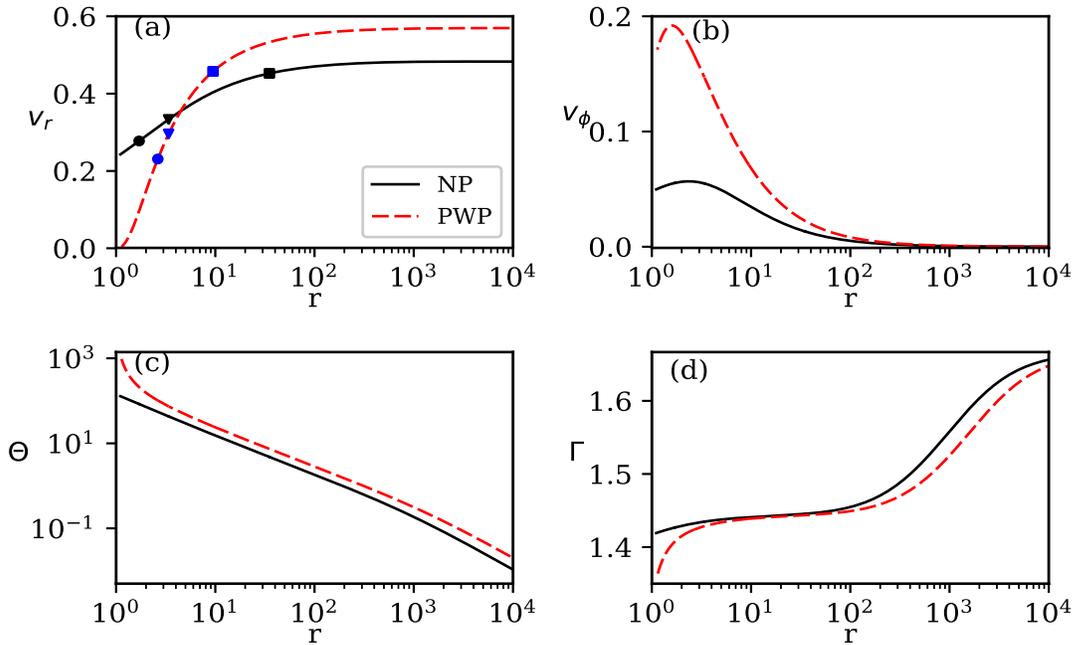} 
\caption{Wind solution in Newtonian potential $\Phi_{\rm NP}$ (solid, black) and
in Paczy\'{n}ski \& Wiita potential $\Phi_{\rm PWP}$ (dashed, red) for $E=1.301$, $L=1.75$ and $\xi=1$.
The flow variables are (a) $\vr$, (b) $\vfi$, 
(c) $\Theta$ and (d) $\Gamma$ versus radius $r$. The slow-point, Alf\'ven point and fast-point are marked
as solid, circle, triangle and square.}
\label{lab:fig8}
\end{figure}

In Figs. (\ref{lab:fig8}) we present the effect of strong gravity. We chose electron-proton or $\xi=1.0$ plasma,
where the flow has the same Bernoulli parameter $E ~(=1.301)$, $\ra~(=3.3859)$ and total angular momentum $L~(=1.75)$. However, we compare
magnetized-wind solutions
expanding in a region described by Newtonian gravitational potential (solid, black) with another one in a region described
by PW
potential (dashed, red). We chose a higher value of $E$, in order to maximize the effect.
Even for the same specific energy and angular momentum, the wind in a region described
by $\Phi_{\rm PWP}$ is faster and hotter than the one in a region described
by a Newtonian gravity. The stronger gravity of a $\Phi_{\rm PWP}$ compresses the plasma around the compact object and produces a higher
temperature flow. The inner boundary condition of the outflow in $\Phi_{\rm \small NP}$, shows that $\vr$ (although sub-slow)
is quite high, while $\vfi$ and $\Theta$ is lower than that around $\Phi_{\rm \small PWP}$. The acceleration
achieved for flows around $\Phi_{\rm \small NP}$ is quite moderate, while that around $\Phi_{\rm \small PWP}$
is significant. If we use lower $E$ then perhaps the $\vr$ at the inner boundary for $\Phi_{\rm \small NP}$ will have proper
value but then the terminal speed would be very low.
One may conclude that, if we use $\Phi_{\rm \small NP}$ to describe the gravity around
a compact object, then, only slow outflows can be obtained.

\vspace{0.0cm}
\begin{figure}
\hspace{0.0cm}
\includegraphics[width=16cm,height=10cm]{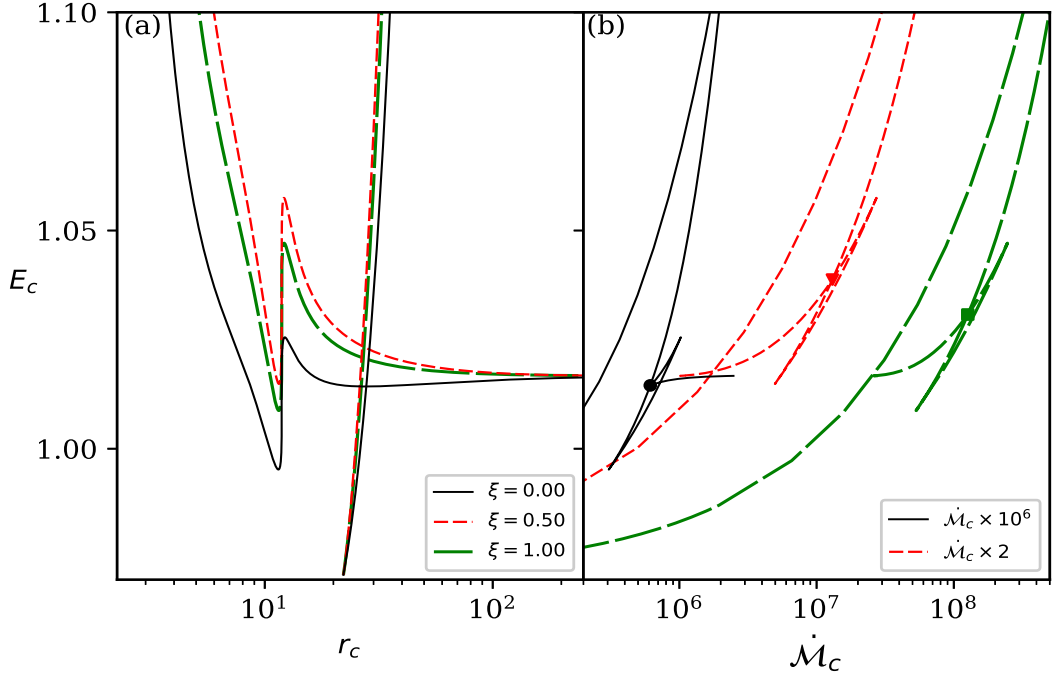} 
\caption{In panel we plot, (a) $E_c$ versus $r_c$ and (b) $E_c$ versus ${\dot {\cal M}}_c$. The curves represent
$\xi=0.0$ (solid, black), $\xi=0.5$ (dashed, red), $\xi=1.0$ (long-dashed, green). All plots are for $L=1.0$.}
\label{lab:fig9}
\end{figure}

In Figs. (\ref{lab:fig9}a \& b), we plot the effect of composition of the flow on its critical point properties.
We plot the Bernoulli parameter as a function of critical radius i.e., $E_c$ versus $r_c$, for fixed value of $L=1.0$ and $\ra=11.85$, but different
values of composition or $\xi=0.0$ (solid, black), $\xi=0.5$ (dashed, red), $\xi=1.0$ (long-dashed, green). Similar to previous
$E_c$ and ${\dot {\cal M}}_c$ plot, the proper wind solution corresponds
to the $E_c$ versus ${\dot {\cal M}}_c$ at the intersection of the X-type slow branch and X-type fast branch (marked as solid circle,
triangle and a square for three values of $\xi$). 

\vspace{0.0cm}
\begin{figure}
\hspace{0.0cm}
\includegraphics[width=16cm,height=10cm]{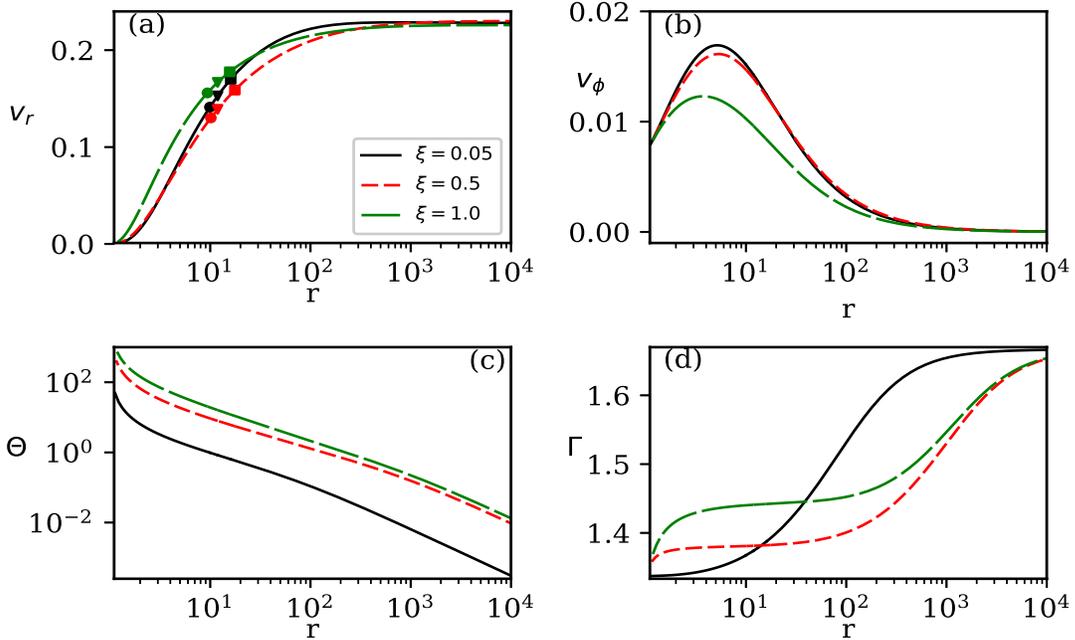} 
\caption{(a) $\vr$, (b) $\vfi$, (c) $\Theta$ and (d) $\Gamma$ as a function of $r$, where each curve represent
$\xi=0.05$ (solid, black), $\xi=0.5$ (dashed, red) and $\xi=1.0$ (long-dashed, green).
All the curves have the same $E=1.03075$ and $L=1.0$.}
\label{lab:fig10}
\end{figure}

In Figs. (\ref{lab:fig10}a---d), we compare $\vr$, $\vfi$, $\Theta$ and $\Gamma$ for flows with the
same $E~(=1.03075)$ and $L~(=1.0)$ but for different composition
$\xi=0.05$ (solid, black), $0.5$ (dashed, red) and $1.0$
(long-dashed, green). All the flow variables like $\vr$, $\vfi$ and $\Theta$ depend on $\xi$.
The $\xi=1.0$ or electron-proton flow has somewhat higher $\vr$ close to the base. The lepton dominated wind
($\xi=0.05$) has higher $\vr$ at some intermediate range of $r$, but finally
for a flow with a composition parameter $\xi=0.5$, the terminal $\vr$ or $\vr^{\rm \small max}=\vr|_{r \rightarrow \infty}$  is higher than the flow with other two combinations, albeit by a small amount. This is not expected in hydrodynamics.
It may be noted that, in hydrodynamics for $r \rightarrow$ large, $h \rightarrow 1$, centrifugal term $\rightarrow 0$, $\Phi(r) \rightarrow 0$,
$\vr^{\rm \small max} \rightarrow \sqrt{2(E-1)}$. In other words, the terminal speeds of winds in hydrodynamics do not depend on
composition, but for MHD winds, even as $r\rightarrow$ large, $h\gsim 1$ because of the presence of the magnetic term and hence
$\vr^{\rm \small max}$ depends on $\xi$. 
The azimuthal velocity $\vfi$ also depends on $\xi$, where an electron-proton flow has the least $\vfi$ distribution when compared with
flows of higher proportion of leptons. However, $\vfi \rightarrow 0$ at $r \rightarrow$ large, therefore the asymptotic value of
$\vfi$ does not depend on $\xi$.
It may be noted that,
the effect of $\xi$ cannot be
studied by attributing some scale factor, as one can clearly see that, the $\vfi$ curves intersect for flows with $\xi=0.5$ and $0.05$.
Similar to all the studies in the hydrodynamic regime in the present case too, the temperature distribution $\Theta(r)$ is highest
for $\xi=1.0$.
Although $\Theta$ is high for $\xi=1.0$ compared to other flows with $\xi<1$, the $\Gamma$ for electron-proton flow
is not the lowest. In fact, at lower values of $r$, $\Gamma_{\xi=0.05} < \Gamma_{\xi=0.5} < \Gamma_{\xi=1}$.
This is because, $\Gamma$ compares the thermal energy of the plasma compared to its inertia, hence
high value of $\Theta$ cannot compensate for higher inertia of an electron-proton flow.
In comparison, a lepton dominate flow ($\xi=0.05$) achieves $\Gamma \rightarrow 4/3$, in spite of starting with a temperature
at least an order of magnitude less that of an electron-proton flow.

\begin{figure}
\hspace{0.0cm}
\includegraphics[width=15cm,height=10cm]{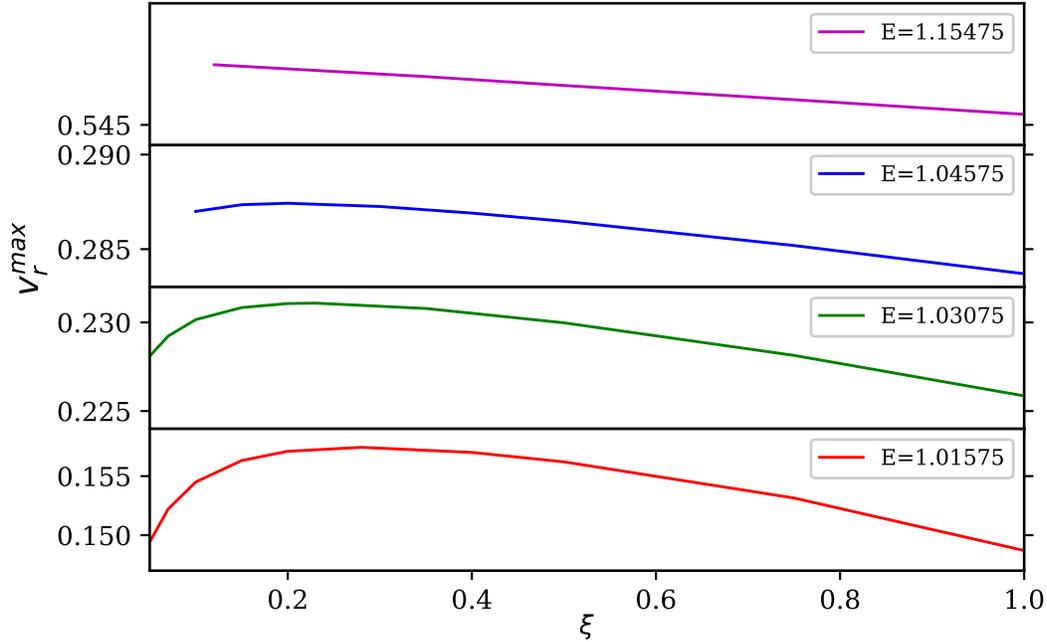} 
\caption{Terminal speed $\vr^{\rm \small max}$ is plotted as a function of $\xi$ from top panel and downwards for
$E=1.15475,~1.04575,~1.03075,\mbox{ and }1.01075$, respectively. $L=1.0$ for all the curves.}
\label{lab:fig11}
\end{figure}

In Figs. (\ref{lab:fig11}) we have plotted $\vr^{\rm \small max}$ as a function of $\xi$,
for various values of energies ($E=1.15475,~1.04575,~1.03075,\mbox{ and }1.01075$) and a given value of $L$.
It is interesting to note that terminal speed distribution for higher to lower energies of the magnetized outflow,
depends on $\xi$. For higher energies, it decreases with $\xi$, but for lower $E$, $\vr^{\rm \small max}$ maximizes at some value
of $\xi$. The maxima depends on $E$.
This is a significantly different result compared to hydrodynamics. This effect arises due to the competition between pressure
gradient and magnetic forces in the equations of motion. 

\vspace{0.0cm}
                   
\section{Discussion and Conclusions}
\label{sec:conclude}

Our main focus is on studying the effect of variable $\Gamma$ EoS and the composition on magnetized wind solutions.
This model is the bedrock over which magnetized jet models were developed later. In this context, it may be noted that there are
of course many models of generation of outflows around compact objects, tailored to address different scenarios. For example,
if the underlying accretion disc is luminous then radiation can drive outflows via scattering processes {\citep{i80,tf96,tf98,ms15,ybl18,
vc19}}
or for cooler gases via the line driven processes \citep{notwy16,no17}. In case the accretion flow is low luminous flow then magnetic,
gas pressure and centrifugal term may drive outflows {\citep{g15,ygnsbb15,bygy16a,bygy16b,bm18}}. In the present paper we are also working in the regime where radiation
is not important. However, our main focus is to study a typical transmagnetosonic outflow solution and what are the parameters these
solutions depend on. In particular, the effect of a variable $\Gamma$ EoS and different composition of the plasma on the
outflow solution has probably not been studied before.

In this paper we have revisited magnetized, wind model using
Paczy\'{n}ski \& Wiita potential, variable $\Gamma$ EoS and for various values of
total angular momentum ($L$), Bernoulli parameter ($E$) and composition ($\xi$) of the flow. $E$ and $L$ are constants of motion.
However, as is the case in hydrodynamics, in MHD too, there can be a plethora of solutions
corresponding to the same set of constants of motion. As has been shown by \citet{b52}, of all the possible solutions
related to a given set of constants of motion, the physical global solution is the one which has the highest entropy and also
happens to be
the transonic one. Similarly in MHD (Figs. \ref{lab:fig2}), it was shown that the solution passing through slow, Alf\'ven and
fast points is the correct solution.
We have found that for a given set of values of $E$ and $\xi$, higher angular momentum
outflows are slower (lesser $\vr$) compared to flows with lower $L$. This is understandable,
since matter which is rotating faster, will not able to possess higher $\vr$.
In particular, for flows with lower $L$, magnetic field may deflect outflow from counter clockwise to a quasi-radial flow relatively close
to the central object (Fig. \ref{lab:fig6}a). In other words the effect of magnetic field cannot be quantified by how much
the flow is accelerated radially, but magnetic field has a very important role in regulating flow angular momentum.
We also show that faster outflow is possible, for flows with higher $E$.
Interestingly, the $\Theta$ differs slightly if $E$ or $L$ is changed, but the change in $\vr$ and $\vfi$ is more significant.
However, $\vr$, $\vfi$ and $\Theta$
distributions are significantly different for different values of $\xi$. 
This is because, if we change the composition of the flow, then we are not only changing its thermal
energy content (which pushes the matter outward), but also inertia of flow.
Therefore, the terminal speed of the outflows at a given value of $E$ and $L$ maximizes at a given value of $\xi$.
Although, the $\xi$ at which the peak of $\vr^{\rm \small max}$ will occur, also depends on $E$. The peak steadily shifts
to higher $\xi$ as $E$ decreases.
Dependence of various flow variables on $\xi$ for MHD flows probably has not been reported before.
In addition, the spectra emitted by such winds should be quite different for different values of $\xi$,
since the different velocity distributions would result in different
density distributions. Moreover, different temperature distributions would strongly determine the processes that would
dominate the emission. We also studied the wind solutions in different
gravitational potentials to show how the compactness of central object affect the outflow solutions.

\appendix

\section{Combination of forces for magnetized outflow} \label{app:appA}
\begin{figure}
\hspace{0.0cm}
\includegraphics[width=15cm,height=10cm]{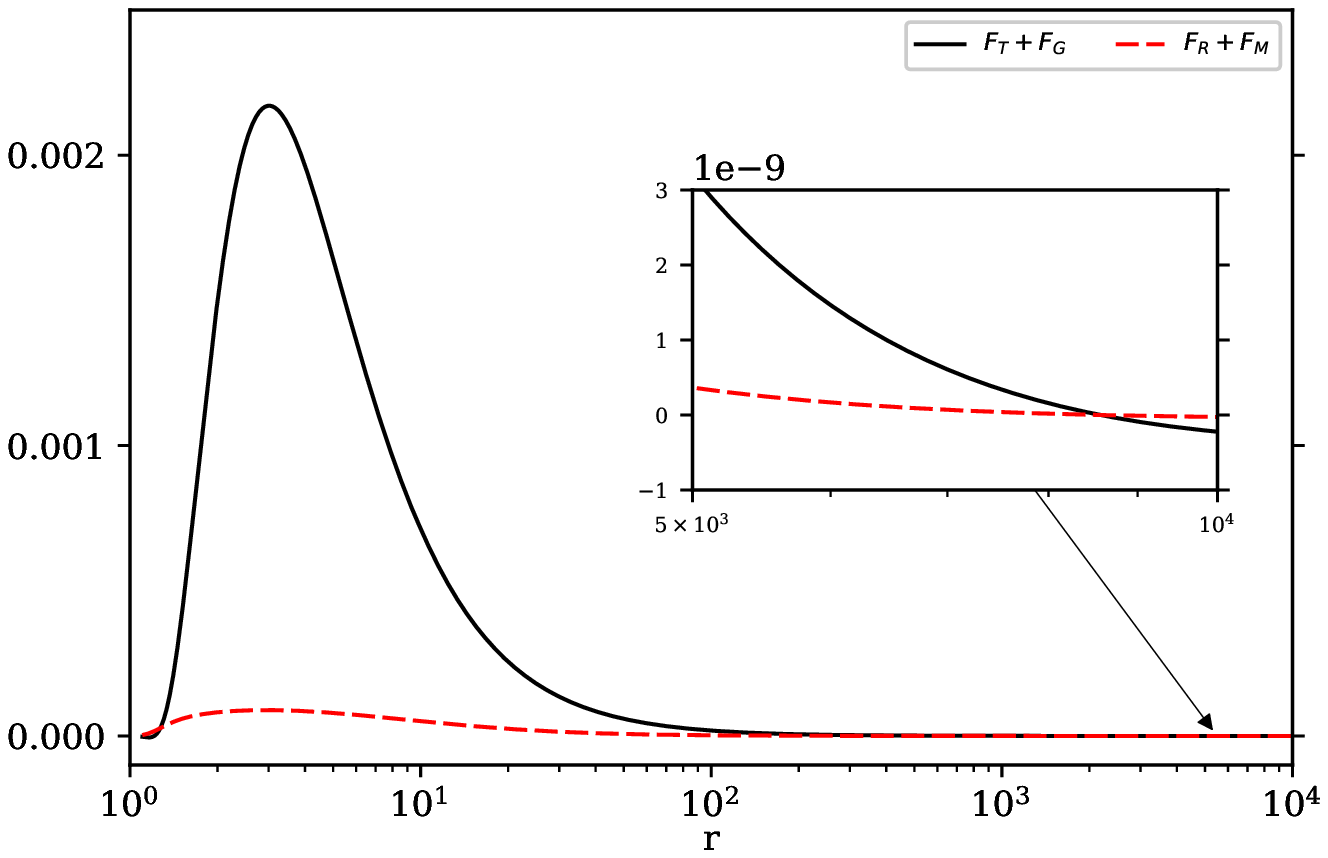} 
\caption{Comparison of combination of forces thermal and gravitational $F_{\rm T}+F_{\rm G}$ (solid, black) and
rotational and magnetic forces $F_{\rm R}+F_{\rm M}$
(dashed, red) along the streamline. The flow parameters are $E=1.03075,~L=1.0$ and $\xi=1.0$ same as Fig. \ref{lab:fig3}.}
\label{lab:figapend}
\end{figure}

{Here we plot the competing combined forces acting on a magnetized wind/outflow. The thermal force is $F_{\rm T}$, rotational
force is $F_{\rm R}$, $F_{\rm G}$ is the gravitational and $F_{\rm M}$ is the magnetic force along the stream line,
as is described in connection with Figs. (\ref{lab:fig3}d). We plot $F_{\rm T}+F_{\rm G}$ (solid, black) and $F_{\rm R}+F_{\rm M}$
(dashed, red) in Fig. \ref{lab:figapend} in order to compare the competing forces. Clearly $F_{\rm R}+F_{\rm M}$ is comparable
$F_{\rm T}+F_{\rm G}$ close to the central object where it is launched, but is larger at very large distances. However,
$F_{\rm T}+F_{\rm G}$ dominates over the other combination very close to the central object till unto a hundred
$\rg$.}

{}
\end{document}